\documentclass[12pt,a4paper,final]{iopart}
\usepackage{iopams}

\expandafter\let\csname equation*\endcsname\relax
\expandafter\let\csname endequation*\endcsname\relax

\usepackage{amsmath}
\usepackage{amsfonts}
\usepackage{amsmath}
\usepackage{amssymb}
\usepackage{amsthm}
\usepackage[utf8]{inputenc}
\usepackage{bm}
\usepackage{cite}

\usepackage[breaklinks=true,colorlinks=true,linkcolor=blue,urlcolor=blue,citecolor=blue]{hyperref}

\newcommand{\defeq}{\mathrel{\mathop:}=}

\newcommand{\U}{\mathcal{U}}
\renewcommand{\H}{\mathcal{H}}

\newtheorem{theorem}{Theorem}

\begin{document}

\title{Fundamental temperature exclusively determines the validity of superstatistics}

\author[cor1]{Sergio Davis$^{1,2}$}
\address{$^1$Research Center on the Intersection in Plasma Physics, Matter and Complexity (P$^2$mc), Comisión Chilena de Energía Nuclear, Casilla 188-D, Santiago, Chile}
\address{$^2$Departamento de Ciencias F\'isicas, Facultad de Ciencias Exactas, Universidad Andres Bello. Sazi\'e 2212, piso 7, 8370136, Santiago, Chile.}
\ead{sergio.davis@cchen.cl}

\author{Constanza Farías$^2$}
\address{$^2$Departamento de Ciencias F\'isicas, Facultad de Ciencias Exactas, Universidad Andres Bello. Sazi\'e 2212, piso 7, 8370136, Santiago, Chile.}

\begin{abstract}
The theory of superstatistics is a generalization of Boltzmann-Gibbs statistical mechanics which admits temperature fluctuations, and generates non-canonical ensembles 
from the distribution function of these fluctuations. Recently, some results have been presented showing that superstatistics is not universally applicable, but several conditions 
on the so-called fundamental inverse temperature function $\beta_F$ must be met by any superstatistical model. In this work we provide a set of neccessary and sufficient conditions 
for a non-equilibrium steady state model to be expressible by superstatistics, showing that $\beta_F$ by itself determines the existence of a superstatistical distribution of temperature.
\end{abstract}

\section{Introduction}

Nonequilibrium steady states are commonly observed in real physical systems, such as plasmas~\cite{Lima2000, Abe2002, Ourabah2015, Ourabah2020b} and self-gravitating systems~\cite{Komatsu2012}, 
as well as in complex, non-physical systems such as financial markets~\cite{Tsallis2003, Denys2016}, social networks~\cite{Deppman2021} and others. 

Among the theoretical generalizations to traditional statistical mechanics aiming to describe these nonequilibrium steady states, Tsallis nonextensive statistics~\cite{Tsallis2009c} and superstatistics~\cite{Beck2003, Beck2004} are arguably the most prevalent in the literature. In particular, superstatistics provides an elegant and compact formalism where the inverse temperature 
$\beta \defeq 1/(k_B T)$ is a random variable with a well-defined probability density. 

Although superstatistics can be postulated in a manner fully consistent with probability theory~\cite{Sattin2006, Davis2018, Sattin2018} and can make use~\cite{Davis2020c} of Jaynes' principle of 
maximum entropy~\cite{Jaynes2003}, it is not compatible with every possible nonequilibrium steady state model. 

\noindent
The problem of establishing the range of validity of superstatistics is then an 
open issue, and consequently, some of us~\cite{Davis2022b} have recently proposed a classification of non-equilibrium steady states where the probability density of microstates $\bm \Gamma$ is of the form
\begin{equation}
P(\bm \Gamma|S) = \rho\big(\H(\bm \Gamma); S\big)
\end{equation}
with $\rho(E; S)$ a non-negative function called the ensemble function and $\H(\bm \Gamma)$ the Hamiltonian of the system. In this classification, superstatistical models occupies only a 
region in the space of all possible steady-state models. In particular, this space of models was divided into two regions, depending on the sign of the inverse temperature covariance
\begin{equation}
\label{eq:Udef}
\U \defeq \big<(\delta \beta_F)^2\big>_S - \big<{\beta_F}'\big>_S
\end{equation}
where $\beta_F$ is the fundamental inverse temperature, defined by
\begin{equation}
\beta_F(E) \defeq -\frac{\partial}{\partial E}\ln \rho(E; S).
\end{equation}
 
According to this classification, models are \emph{supercanonical} when $\U > 0$, and \emph{subcanonical} for $\U < 0$, with the canonical ensemble (representing thermal equilibrium) 
corresponding to\;\;$\U = 0$. Superstatistical models are supercanonical, as in that case $\U$ coincides with the variance of the inverse temperature, that is,
\begin{equation}
\U = \big<(\delta \beta)^2\big>_S \geq 0.
\end{equation}

In this work, we present a set of neccessary and sufficient conditions for the validity of superstatistics, expressed only in terms of the fundamental inverse temperature function $\beta_F$ 
and its derivatives.

The remainder of this work is organized as follows. In Section \ref{sec:superstat}, we present a brief introduction to the superstatistical framework, together with some known neccesary conditions for 
its validity. Next, in Section \ref{sec:theorem} we state the main result of this work (whose proof is presented in the \ref{sec:appendix}), while in Section \ref{sec:recurrence} we provide some 
expectation identities that lead to shortcuts in the computation of moments and cumulants. In Section \ref{sec:examples} we provide some concrete examples of the application of our results and, 
finally, we close our discussion with some concluding remarks in Section \ref{sec:concluding}.

\section{The superstatistical framework}
\label{sec:superstat}

Traditional, Boltzmann-Gibbs statistical mechanics is based on the canonical ensemble, where the probability (density) of observing a microstate $\bm \Gamma$ at a temperature $T$ is given by 
\begin{equation}
\label{eq:canonical}
P(\bm \Gamma|\beta) = \frac{\exp\big(-\beta\H(\bm \Gamma)\big)}{Z(\beta)},
\end{equation}
with $\beta \defeq 1/(k_B T)$ the inverse temperature and $Z(\beta) \defeq \int d\bm{\Gamma}\exp\big(-\beta\H(\bm \Gamma)\big)$ the partition function. Typically, $\bm \Gamma$ is a point in the 
phase space of the system, for instance, $\bm \Gamma = (\bm{r}_1, \ldots, \bm{r}_N, \bm{p}_1, \ldots, \bm{p}_N)$ for a system of $N$ particles, the $i$-th particle having position $\bm{r}_i$ 
and momentum $\bm{p}_i$.

Superstatistics takes this canonical ensemble and extends it, postulating that the inverse temperature $\beta$ is no longer fixed but an additional degree of freedom of the system. Therefore, 
the canonical distribution in \eqref{eq:canonical} is now replaced by a joint distribution of $\bm \Gamma$ and $\beta$, namely
\begin{equation}
P(\bm \Gamma, \beta|S) = P(\bm \Gamma|\beta, S) P(\beta|S) = P(\beta|S) \frac{\exp\big(-\beta\H(\bm \Gamma)\big)}{Z(\beta)}.
\end{equation}

As, in practice, we are only interested in the marginal probability (density) of microstates $P(\bm \Gamma|S)$, we integrate the joint distribution over $\beta$ and we obtain
\begin{equation}
\label{eq:super_micro}
P(\bm \Gamma|S) = \int_0^\infty d\beta\, P(\beta|S)\frac{\exp\big(-\beta\H(\bm \Gamma)\big)}{Z(\beta)},
\end{equation}
which, depending on the functional form of $P(\beta|S)$, can lead to an ensemble quite different from the canonical. We clearly see that $P(\bm \Gamma|S)$ depends on $\bm \Gamma$ only 
through the Hamiltonian $\H(\bm \Gamma)$, thus we can define a non-negative function $\rho(E; S)$, called the ensemble function associated to $S$, such that
\begin{equation}
\label{eq:rho}
P(\bm \Gamma|S) = \rho\big(\H(\bm \Gamma); S\big).
\end{equation}

\noindent
From comparison of \eqref{eq:rho} and \eqref{eq:super_micro} we readily see that
\begin{equation}
\rho(E; S) = \int_0^\infty d\beta\;f(\beta; S)\exp(-\beta E)
\end{equation}
that is, $\rho(E; S)$ is the Laplace transform of a new function,
\begin{equation}
f(\beta; S) \defeq \frac{P(\beta|S)}{Z(\beta)},
\end{equation}
that we will call the superstatistical weight function.

Often the formalism of superstatistics is written in terms of values of energy instead of the microstates. For instance, the joint probability density of energy and inverse temperature is
\begin{equation}
\label{eq:prob_E_beta}
\begin{split}
P(E, \beta|S) & = \int d\bm{\Gamma} P(E|\bm \Gamma, \beta)P(\bm \Gamma, \beta|S) \\
& = \int d\bm{\Gamma} \delta\big(\H(\bm \Gamma)-E\big)P(\bm \Gamma, \beta|S) \\
& = \exp(-\beta E)f(\beta; S)\Omega(E),
\end{split}
\end{equation}
where $\Omega(E) = \int d\bm{\Gamma}\,\delta\big(\H(\bm \Gamma)-E\big)$ is the density of states. Similarly, the marginal distribution of energy is given by
\begin{equation}
\label{eq:prob_E}
P(E|S) = \int_0^\infty d\beta P(E, \beta|S) = \int_0^\infty d\beta \exp(-\beta E)f(\beta; S)\Omega(E) = \rho(E; S)\Omega(E).
\end{equation}
 
From \eqref{eq:prob_E} and \eqref{eq:prob_E_beta} we can obtain the probability density of inverse temperature given an observed value of energy $E$, namely
\begin{equation}
\label{eq:prob_beta_given_E}
P(\beta|E, S) = \frac{P(E, \beta|S)}{P(E|S)} = \frac{\exp(-\beta E)f(\beta; S)}{\rho(E; S)}.
\end{equation}

This quantity will be central to our analysis in the following sections. In particular, let us compute the expected value of $\beta$ given $E$, that is,
\begin{equation}
\label{eq:beta_cond}
\big<\beta\big>_{E, S} = \int_0^\infty d\beta P(\beta|E, S)\beta.
\end{equation}

\noindent
Replacing \eqref{eq:prob_beta_given_E} into \eqref{eq:beta_cond} we have
\begin{equation}
\begin{split}
\big<\beta\big>_{E, S} & = \frac{1}{\rho(E; S)}\int_0^\infty d\beta\,f(\beta; S)\exp(-\beta E)\beta \\
& = -\frac{1}{\rho(E; S)}\frac{\partial}{\partial E}\int_0^\infty d\beta\,f(\beta; S)\exp(-\beta E),
\end{split}
\end{equation}
that is,
\begin{equation}
\label{eq:beta_betaF}
\big<\beta\big>_{E, S} = \beta_F(E; S).
\end{equation}

Two neccessary conditions for the validity of superstatistics are already known, and they involve the sign of $\beta_F$ and its derivative ${\beta_F}'$. First, since $\beta_F$ is the expected value of a 
non-negative quantity $\beta$ according to \eqref{eq:beta_betaF}, we have
\begin{equation}
\beta_F(E; S) \geq 0\;\;\text{for}\;E \geq 0.
\end{equation}

\noindent 
Second, differentiating both sides of \eqref{eq:beta_betaF} written as
\begin{equation}
\big<\beta\big>_{E, S} = -\frac{1}{\rho(E; S)}\frac{\partial \rho(E; S)}{\partial E}
\end{equation}
we have
\begin{equation}
\frac{\partial}{\partial E}\big<\beta\big>_{E, S} = \frac{1}{\rho(E; S)^2}\left(\frac{\partial \rho(E; S)}{\partial E}\right)^2 - \frac{1}{\rho(E; S)}\frac{\partial^2 \rho(E; S)}{\partial E^2}
\end{equation}
hence
\begin{equation}
\label{eq:betaF_der}
{\beta_F}'(E; S) = \beta_F(E; S)^2 - \big<\beta^2\big>_{E, S} = -\big<(\delta \beta)^2\big>_{E, S},
\end{equation}
and because the variance in the right-hand side is non-negative, we have the inequality
\begin{equation}
\label{eq:second_ineq}
{\beta_F}'(E; S) \leq 0\;\;\text{for}\;E \geq 0.
\end{equation}

Note that replacing \eqref{eq:second_ineq} into \eqref{eq:Udef} implies the neccessary condition $\U \geq 0$. The question remains about the existence of further neccessary conditions involving 
the second or higher-order derivatives of $\beta_F$.

\section{Fundamental temperature determines the superstatistical class of models}
\label{sec:theorem}

In this section, we will show that all positive moments of $\beta$ given $E$ for a superstatistical system can be directly computed using only $\beta_F$ and its derivatives. Therefore, the function 
$\beta_F(E; S)$ by itself determines the existence of $P(\beta|E, S)$ in \eqref{eq:prob_beta_given_E}. In order to show why this is true, we first obtain the general expression for the $n$-th moment 
in terms of $\rho(E; S)$,
\begin{equation}
\big<\beta^n\big>_{E, S} = \frac{1}{\rho(E; S)}\int_0^\infty d\beta\,f(\beta; S)\exp(-\beta)\beta^n = \frac{(-1)^n}{\rho(E; S)}\frac{\partial^n \rho(E; S)}{\partial E^n}
\end{equation}
of which \eqref{eq:beta_betaF} is the special case with $n = 1$. Now we use Fa\`{a} di Bruno's formula for the $n$-th derivative of a composite function,
\begin{equation}
\frac{\partial^n}{\partial E^n}f\big(g(E)\big) = \sum_{k=1}^n f^{(k)}\big(g(E)\big) B_{n, k}\big(g'(E), g''(E), g'''(E), \ldots, g^{(n-k+1)}(E)\big)
\end{equation}
letting $f(z) = \exp(z)$ and $g(E) = \ln \rho(E; S)$. Here $B_{n, k}$ are the partial exponential Bell polynomials~\cite{Bell1927, Riordan2014}, defined by
\begin{equation}
B_{n, k}(x_1, x_2, \ldots, x_{n-k+1}) \defeq \sum \frac{n!}{j_1!j_2!\ldots j_{n-k+1}!}\left(\frac{x_1}{1!}\right)^{j_1}\left(\frac{x_2}{2!}\right)^{j_2} \ldots \left(\frac{x_{n-k+1}}{(n-k+1)!}\right)^{j_{n-k+1}}.
\end{equation}

\noindent
Because $f^{(k)}(z) = f(z)$ for $f(z) = \exp(z)$, we obtain
\begin{equation}
\big<\beta^n\big>_{E, S} = (-1)^n B_n\left(-\beta_F, -{\beta_F}', -{\beta_F}'', \ldots, -{\beta_F}^{(n-1)}\right),
\end{equation}
with $B_n$ the $n$-th complete exponential Bell polynomial, given by
\begin{equation}
B_n(x_1, \ldots, x_n) \defeq \sum_{k=1}^n B_{n, k}(x_1, x_2, \ldots, x_{n-k+1}),
\end{equation}
and by virtue of the properties of $B_{n, k}$, we have finally
\begin{equation}
\label{eq:moments_explicit}
\big<\beta^n\big>_{E, S} = B_n\left(\beta_F, -{\beta_F}', {\beta_F}'', \ldots, (-1)^{n-1} {\beta_F}^{(n-1)}\right).
\end{equation}

We thus see that $\beta_F$ and its derivatives determine the set of all the moments of $P(\beta|E, S)$, fixing in turn the distribution itself. Moreover, due to the fact that 
\[P(\beta|E, S) \propto \exp(-\beta E)f(\beta; S),\] it follows that $f(\beta; S)$ is also uniquely determined by $\beta_F$. As an example of \eqref{eq:moments_explicit}, the first four moments of 
$P(\beta|E, S)$ evaluate to
\begin{subequations}
\begin{align}
\label{eq:mom1}
\big<\beta\big>_{E, S} & = \beta_F, \\
\label{eq:mom2}
\big<\beta^2\big>_{E, S} & = (\beta_F)^2 -{\beta_F}', \\
\big<\beta^3\big>_{E, S} & = (\beta_F)^3 - 3\beta_F{\beta_F}' + {\beta_F}'', \\
\big<\beta^4\big>_{E, S} & = (\beta_F)^4 - 6(\beta_F)^2{\beta_F}' + 3({\beta_F}')^2 + 4\beta_F {\beta_F}'' - {\beta_F}''',
\end{align}
\end{subequations}
where \eqref{eq:mom1} and \eqref{eq:mom2} agree with \eqref{eq:beta_betaF} and \eqref{eq:betaF_der}, respectively.

We can understand the meaning of \eqref{eq:moments_explicit} by recalling the concept of cumulants in probability theory~\cite{Grimmett2014}. The cumulants $\kappa_1, \kappa_2, \kappa_3, \ldots$ 
are similar to the moments of a probability distribution, but are defined through the cumulant-generating function~\cite{Kendall1958},
\begin{equation}
\label{eq:cumulant_series}
\ln M_\beta(t; E, S) \defeq \sum_{n=1}^\infty \frac{t^n}{n!}\,\kappa_n(E; S),
\end{equation}
where $M_\beta(t; E, S)$ is the moment-generating function for $P(\beta|E, S)$, in turn defined by
\begin{equation}
\label{eq:mbeta_def}
M_\beta(t; E, S) \defeq \big<\exp(t\beta)\big>_{E, S}.
\end{equation}

\noindent
Replacing \eqref{eq:prob_beta_given_E} into \eqref{eq:mbeta_def} we obtain
\begin{equation}
\begin{split}
M_\beta(t; E, S) & = \int_0^\infty d\beta\;P(\beta|E, S)\exp(t\beta) \\
& = \int_0^\infty d\beta\,\frac{f(\beta; S)\exp\big(-\beta [E-t]\big)}{\rho(E; S)} = \frac{\rho(E-t; S)}{\rho(E; S)}
\end{split}
\end{equation}
therefore, from the Taylor expansion of $\ln \rho(E-t; S)$ around $t = 0$ we have
\begin{equation}
\begin{split}
\ln M_\beta(t; E, S) & = \ln \rho(E-t; S) - \ln \rho(E; S) \\
& = \sum_{n=0}^\infty (-1)^n \frac{t^n}{n!}\left[\frac{\partial^n}{\partial E^n}\ln \rho(E; S)\right] -\ln \rho(E; S) \\
& = \sum_{n=1}^\infty (-1)^{n-1}\frac{t^n}{n!}\beta^{(n-1)}_F(E; S).
\end{split}
\end{equation}

By comparing with the power series in \eqref{eq:cumulant_series} term by term we see that the $n$-th cumulant is given in terms of the $(n-1)$-th derivative of $\beta_F$ by
\begin{equation}
\label{eq:cumulants}
\kappa_n(E; S) = (-1)^{n-1}\beta^{(n-1)}_F(E; S).
\end{equation}

These results, in particular the formulas \eqref{eq:moments_explicit} and \eqref{eq:cumulants}, motivate the following theorem regarding the sign of the $n$-th derivative of $\beta_F$, theorem 
which is proved in the \ref{sec:appendix}.

\begin{theorem}
\label{thm:theorem1}
A steady-state model $S$ having fundamental inverse temperature $\beta_F$ is a superstatistical model (including the canonical case) if and only if
\begin{equation}
\label{eq:the_condition}
(-1)^n \beta_F^{(n)}(E; S) \geq 0
\end{equation}
holds for all integer $n \geq 0$.
\end{theorem}

In other words, in superstatistics all even derivatives of $\beta_F$ must be positive or zero, while all odd derivatives must be negative or zero. Condition \eqref{eq:the_condition} is both 
a neccessary and sufficient condition for superstatistics to be valid, the latter allows us to use the set of inequalities in \eqref{eq:the_condition} as an alternative definition of a 
superstatistical model without explicitly introducing an inverse temperature distribution.

Theorem \ref{thm:theorem1} also implies that the cumulants $\kappa_n(E; S)$ of $P(\beta|E, S)$ are all non-negative, while a further consequence of \eqref{eq:cumulants} is implied by Marcinkiewicz's theorem~\cite{Marcinkiewicz1939}, stating that no probability distribution can have a cumulant-generating function that is a polynomial 
of degree greater than 2. This means that
\begin{equation}
\big|\kappa_n(E; S)\big| > 0\;\;\text{for}\;n \geq 3,
\end{equation}
and, consequently from \eqref{eq:cumulants},
\begin{equation}
\big|\beta_F^{(n)}(E; S)\big| > 0\;\;\text{for}\;n \geq 2
\end{equation}
unless $\beta_F(E; S)$ is the constant function (i.e, when we are in the canonical ensemble). This result tells us that a superstatistical $\beta_F(E; S)$ must be infinitely differentiable and thus 
cannot be a polynomial of any degree in $E$. This observation leads to a stronger variant of theorem \ref{thm:theorem1}.

\begin{theorem}
\label{thm:theorem2}
A steady-state model $S$ having fundamental inverse temperature $\beta_F$ is a non-canonical superstatistical model if and only if
\begin{equation}
\beta_F(E; S) \geq 0,
\end{equation}
and
\begin{equation}
\label{eq:the_condition_strong}
(-1)^n \beta_F^{(n)}(E; S) > 0
\end{equation}
hold for all integer $n \geq 1$.
\end{theorem}

\section{Cumulants using recurrence relations and differential equations}
\label{sec:recurrence}

A sometimes simpler technique to deal with the cumulants and moments of $P(\beta|E, S)$ is the use of expectation identities, in particular the one known as the fluctuation-dissipation 
theorem~\cite{Davis2016c}, which is the identity
\begin{equation}
\frac{\partial}{\partial E}\big<\omega\big>_{E, S} = \left<\omega\frac{\partial}{\partial E}\ln P(\beta|E, S)\right>_{E, S}
\end{equation}
valid for any function $\omega(\beta)$. Replacing $P(\beta|E, S)$ according to \eqref{eq:prob_beta_given_E}, it reduces to
\begin{equation}
\label{eq:fdt}
\frac{\partial}{\partial E}\big<\omega\big>_{E, S} = \big<\omega\big>_{E, S}\,\beta_F(E; S) - \big<\beta\omega\big>_{E, S}.
\end{equation}

Under the choice $\omega(\beta) = \beta^n$ with integer $n$, we obtain a recurrence relation for the moments, namely
\begin{equation}
\big<\beta^{n+1}\big>_{E, S} = \left(\beta_F(E; S)-\frac{\partial}{\partial E}\right)\big<\beta^n\big>_{E, S}.
\end{equation}

We can either use this identity on its own to compute all the positive moments starting from $\beta_F$ without the use of \eqref{eq:moments_explicit}, or use the choice 
$\omega(\beta) = \exp(t\beta)$ into \eqref{eq:fdt}, obtaining a differential equation for the moment-generating function,
\begin{equation}
\frac{\partial}{\partial t}M_\beta(t; E, S) = \left(\beta_F(E; S)-\frac{\partial}{\partial E}\right)M_\beta(t; E, S).
\end{equation}

Dividing both sides by $M_\beta(t; E, S)$ which is never zero, we arrive at an even simpler differential equation for the cumulant-generating function,
\begin{equation}
\label{eq:cgf_partial}
\left(\frac{\partial}{\partial t} + \frac{\partial}{\partial E}\right)\ln M_\beta(t; E, S) = \beta_F(E; S).
\end{equation}

\section{Examples}
\label{sec:examples}

In this section, we will explore several examples of the application of theorems \ref{thm:theorem1} and \ref{thm:theorem2}.

\subsection{The $q$-canonical ensemble}

First, let us consider the $q$-canonical ensemble of Tsallis nonextensive statistics, whose ensemble function is
\begin{equation}
\rho(E; \beta_0, q) = \frac{1}{Z_q(\beta_0)}\Big[1+(q-1)\beta_0 E\Big]_+^{\frac{1}{1-q}}.
\end{equation}

\noindent
The corresponding fundamental inverse temperature function is given by
\begin{equation}
\beta_F(E; \beta_0, q) = \frac{\beta_0}{1+(q-1)\beta_0 E}
\end{equation}
and is such that ${\beta_F}'$ can be conveniently written in terms of $\beta_F$ itself,
\begin{equation}
\label{eq:qcanon_deriv}
{\beta_F}'(E; \beta_0, q) = -(q-1){\beta_F}^2(E; \beta_0, q).
\end{equation}

\noindent
Further differentiation of \eqref{eq:qcanon_deriv} gives, for the higher-order derivatives of $\beta_F$,
\begin{equation}
(-1)^n {\beta_F}^{(n)} = (q-1)^n (n!)\beta_F(E: \beta_0, q)^{n+1},
\end{equation}
thus by comparison with \eqref{eq:the_condition} we see that a superstatistical representation exists if and only if $q \geq 1$.

\subsection{The Gaussian ensemble}

On the other hand, for the Gaussian ensemble~\cite{Challa1988, Challa1988a, Johal2003, Suzuki2022} we have
\begin{equation}
\rho(E; A, \varepsilon) = \frac{1}{\eta_A(\varepsilon)}\exp\left(-A(E-\varepsilon)^2\right)
\end{equation}
with fundamental inverse temperature
\begin{equation}
\beta_F(E; A, \varepsilon) = 2A(E-\varepsilon).
\end{equation}

Here we see that $\beta_F$ can be negative for $E < \varepsilon$, and, moreover, ${\beta_F}' = 2A > 0$, thus there is no superstatistical 
representation for the Gaussian ensemble with $A > 0$. Furthermore, in this case $\beta_F$ is a polynomial, thus superstatistics is ruled out by theorem \ref{thm:theorem2}.

\subsection{A simple correction to the canonical ensemble}

Consider now the model with fundamental inverse temperature
\begin{equation}
\label{eq:example_betaF}
\beta_F(E; \beta_0) = \beta_0 + \frac{1}{E}.
\end{equation}
Its $n$-th derivative for $n \geq 1$ is given by
\begin{equation}
\label{eq:example_nder}
{\beta_F}^{(n)}(E; \beta_0) = \frac{(-1)^n \;n!}{E^{n+1}},
\end{equation}
so the model must have a superstatistical representation. Directly using \eqref{eq:cgf_partial}, we obtain the cumulant-generating function,
\begin{equation}
\ln M_\beta(t; E, \beta_0) = \beta_0 E + \ln E + C(E-t)
\end{equation}
where $C(z)$ is a function to be determined. Imposing that $\ln M_\beta(0; E, \beta_0) = 0$ we have
\begin{equation}
C(z) = -\beta_0 z - \ln z
\end{equation}
therefore
\begin{equation}
\label{eq:example_cgf}
\ln M_\beta(t; E, \beta_0) = \beta_0 t + \ln E - \ln (E-t) = \left(\beta_0 + \frac{1}{E}\right)t + \sum_{n=2}^\infty \frac{t^n}{n!} \frac{(n-1)!}{E^n}
\end{equation}
in other words,
\begin{equation}
\kappa_n(E; \beta_0) = \begin{cases}
\displaystyle\beta_0 + \frac{1}{E}\;\;\text{for}\;n = 1, \\[10pt]
\displaystyle\frac{(n-1)!}{E^n}\;\;\text{for}\;n > 1,
\end{cases}
\end{equation}
in agreement with \eqref{eq:cumulants} and \eqref{eq:example_nder}. In fact, the ensemble function corresponding to \eqref{eq:example_betaF} is
\begin{equation}
\rho(E; \beta_0) = \frac{\exp(-\beta_0 E)}{\zeta(\beta_0)E},
\end{equation}
which is the Laplace transform of
\begin{equation}
f(\beta; \beta_0) = \frac{\Theta(\beta-\beta_0)}{\zeta(\beta_0)}
\end{equation}
so from \eqref{eq:prob_beta_given_E} we can verify that the conditional density
\begin{equation}
P(\beta|E, \beta_0) = \exp(\beta_0 E) E\exp\big(-\beta E\big)\Theta(\beta-\beta_0)
\end{equation}
is correctly normalized and has the moment-generating function given by
\begin{equation}
M_\beta(t; E, \beta_0) = \int_0^\infty d\beta P(\beta|E, \beta_0)\exp(\beta t) = \frac{E}{E-t}\exp(\beta_0 t),
\end{equation}
agreeing with \eqref{eq:example_cgf}.

\section{Concluding remarks}
\label{sec:concluding}

We have established two theorems, both providing neccessary and sufficient conditions for the validity of superstatistics. The stronger of the two, theorem \ref{thm:theorem2}, excludes the 
trivial case of the canonical ensemble, where $\beta_F$ is the constant function. Explicit formulas for the moments and cumulants of the conditional distribution $P(\beta|E, S)$ are given, 
expressed exclusively in terms of $\beta_F$ and its derivatives. A corollary of theorem \ref{thm:theorem2} is that the fundamental inverse temperature functions of non-canonical superstatistical 
models are infinitely differentiable, and thus cannot be polynomials of any order.

\section*{Acknowledgments}

The authors are grateful to Daniel Pons for fruitful discussions on the subject of completely monotone functions.
We gratefully acknowledge funding from ANID FONDECYT 1220651 grant (SD, CF) and Beca ANID Doctorado Nacional/(2021) - 21210658 (CF). 

\appendix
\section{Proof of theorem 1}
\label{sec:appendix}

Let us first recall the definition of a completely monotone function and a Bernstein function, following the book by Schilling, Song and Vondracek~\cite{Schilling2010}. 
Letting $\mathcal{CM}$ be the set of all completely monotone functions, from Definition 1.3 of Ref.~\cite{Schilling2010} we have that, for a function $F(x)$ with $x > 0$ and integer $n$,
\begin{equation}
\label{eq:monotone_deriv}
F \in \mathcal{CM}\;\;\text{if and only if}\;\;(-1)^n F^{(n)}(x) \geq 0\;\;\text{for}\;n \geq 0.
\end{equation}

According to Theorem 1.4 of Ref.\cite{Schilling2010} (Bernstein's theorem), a function $F$ is completely monotone if and only if it can be expressed as the Laplace transform of another, 
non-negative function $G$, that is, 
\begin{equation}
\label{eq:monotone_laplace}
F \in \mathcal{CM}\;\;\text{if and only if}\;\;F(x) = \int_0^\infty ds\, G(s)\exp(-sx)\;\;\text{with}\;G(s) \geq 0.
\end{equation}

On the other hand, from Definition 3.1 of Ref.~\cite{Schilling2010}, $H(x)$ is a Bernstein function if and only if $H(x) \geq 0$ and $H'(x)$ is completely monotone. Furthermore, denoting 
by $\mathcal{BF}$ the set of all Bernstein functions, from Theorem 3.6 of Ref.~\cite{Schilling2010} we have that the composite function $F(H(x)) \in \mathcal{CM}$ if and only if 
$F \in \mathcal{CM}$ and $H \in \mathcal{BF}$.

\begin{proof}[Proof of theorem 1]
Clearly, from \eqref{eq:monotone_deriv}, the condition \eqref{eq:the_condition} is equivalent to the assertion that $\beta_F(E; S)$ is completely monotone. On the other hand, the assertion that 
$\rho(E; S)$ is a superstatistical model is equivalent, because of \eqref{eq:monotone_laplace}, to the assertion that $\rho(E; S)$ is completely monotone. Therefore, the proof of our main theorem 
reduces to proving the proposition
\begin{equation}
\label{eq:lemma}
\beta_F(E; S) \in \mathcal{CM}\;\;\text{if and only if}\;\;\rho(E; S) \in \mathcal{CM}.
\end{equation}

\noindent
The proof of \eqref{eq:lemma} proceeds by constructing the function
\begin{equation}
H(E) \defeq \ln \rho(E_0; S) - \ln \rho(E; S),
\end{equation}
where $E_0$ is a reference energy, and $H'(E) = \beta_F(E; S)$. By choosing the completely monotone function $F(z) = \exp(-z)$ we see that
\begin{equation}
F(H(E)) = \exp\big(\ln \rho(E; S)-\ln \rho(E_0; S)\big) = \frac{\rho(E; S)}{\rho(E_0; S)}.
\end{equation}

Now, because $\rho(E; S) \geq 0$ for any steady-state model, $F(H(E))$ is completely monotone if and only if $\rho(E; S)$ is completely monotone. Therefore, we have that $\rho(E; S)$ is completely 
monotone if and only if $H(E)$ is a Bernstein function, that is, if and only if $\beta_F(E; S)$ is completely monotone, which is \eqref{eq:lemma}.

\end{proof}

\section*{References}

\bibliography{supcond}
\bibliographystyle{unsrt}

\end{document}